# Topology of the correlation networks among major currencies using hierarchical structure methods

Mustafa Keskin[a, 1], Bayram Deviren[b] and Yusuf Kocakaplan[c]

[a] *Department of Physics, Erciyes University, 38039 Kayseri, Turkey*
[b] *Department of Physics, Nevsehir University, 50300 Nevsehir, Turkey*
[c] *Institute of Science, Erciyes University, 38039 Kayseri, Turkey*

**Abstract**

We studied the topology of correlation networks among 34 major currencies using the concept of a minimal spanning tree and hierarchical tree for the full years of 2007-2008 when major economic turbulence occurred. We used the USD (US Dollar) and the TL (Turkish Lira) as numeraires in which the USD was the major currency and the TL was the minor currency. We derived a hierarchical organization and constructed minimal spanning trees (MSTs) and hierarchical trees (HTs) for the full years of 2007, 2008 and for the 2007-2008 periods. We performed a technique to associate a value of reliability to the links of MSTs and HTs by using bootstrap replicas of data. We also used the average linkage cluster analysis for obtaining the hierarchical trees in the case of the TL as the numeraire. These trees are useful tools for understanding and detecting the global structure, taxonomy and hierarchy in financial data. We illustrated how the minimal spanning trees and their related hierarchical trees developed over a period of time. From these trees we identified different clusters of currencies according to their proximity and economic ties. The clustered structure of the currencies and the key currency in each cluster were obtained and we found that the clusters matched nicely with the geographical regions of corresponding countries in the world such as Asia or Europe. As expected the key currencies were generally those showing major economic activity.

## 1. Introduction

Currency markets which represent the most liquid and largest financial market are extremely important because they perform daily transactions totaling trillions of US dollars, exceeding the yearly gross domestic product (GDP) of most countries [1]. There is no doubt that the value of a currency is extremely important because it is expected to reflect the entire economic status of the country, and the foreign exchange rate is considered to be a measure of economic balance between the two countries. From the physics perspective, foreign exchange markets are typical open systems having interactions with all kinds of financial information around the world including price changes in other markets. The average transaction intervals of foreign exchange markets are typically about 10 s, and it is not clear how the market correlates with the huge scale of information of a whole country or from economic blocks [2]. It is important to investigate the interaction of currencies using the high precision data of foreign exchange markets in order to empirically establish the relations between microscopic market fluctuations and macroscopic economic states.

---
[1] Corresponding author.
Tel: + 90 (352) 4374938#33105; Fax: + 90 (352) 4374931
E-mail address: keskin@erciyes.edu.tr (M. Keskin)

One of the problems in foreign exchange research is that currencies are priced against each other so no independent numeraire exists. Any currency chosen as a numeraire will be excluded from the results, yet its intrinsic patterns can indirectly affect overall patterns. There is no standard solution to this issue or a standard numeraire candidate. Gold was considered, but rejected due to its high volatility. This is an important problem as different numeraires will give different results if strong multidimensional cross-correlations are present. Different bases can also generate different tree structures. The inclusion or exclusion of currencies from the sample can also give different results. This implies that we should take all major currencies and improper emphasis should not be placed on any particular MST result. Resulting robustness should also be checked by comparison with other methods or samples [3].

In the present paper, using the concept of minimal spanning tree (MST) and hierarchical tree (HT), we analyzed the topology of thirty-four major currencies, which were generally free floating and had either market dominance or represented a region covering the data period from January 4, 2007 to December 31, 2008. We choose this period because major economic turbulence occurred during this period. For example, at the beginning of 2007 (02.01.2007), the parity of the USD- TL was about 1.406, falling to 1.215 on 15.10.2007, and dreaching its lowest value at the end of the year (31.12.2007) at 1.1703. At the beginning of 2008 (02.01.2008), the parity was about 1.1708, but it increased to 1.5394 due to economic turbulences on 31.12.2008. Data were provided from the Pacific Exchange Rate Service [4]. We use the USD and TL as numeraires. The USD was a major currency, and the TL was a minor currency. We also performed the bootstrap technique [5] to associate a value of reliability to the links of MSTs and HTs. Finally, we used the average linkage cluster analysis [5, 6] for obtaining the hierarchical trees in the case of the TL as the numeraire.

The MST and HT provided a useful guide for determining underlying economic or regional causal connections in individual currencies. They also show the interconnection among currencies, detecting clusters and taxonomic relations in foreign exchange markets. These trees, introduced by Mantegna [7], and Mantegna and Stanley [8], have been applied to analyze currency markets [2, 9-10], especially to find clustered structure of currencies and the key currency in each cluster [2, 3] and to resolve contagion in a currency crisis [9, 10]. These trees are also used to study the clustering behavior of individual stocks within a single country, for example, usually the USA [11-15], Korea and the USA [16], Italy [17, 18], Greece [19], and also in Turkey [20]. The concept of the MST and the HT is also used to examine the extent and evolution of interdependence between world equity markets [21, 22], European equity markets [23] and commodity markets [24]. Finally, we should also mention that a variety of dynamic MST analysis has also been developed and applied to the time-varying behavior of stocks in Refs. [25-30].

The outline of the remaining part of this paper is organized as follows. In Section 2, we briefly introduce the MST, ultrametric distance and the HT constructed from the Pearson correlation coefficient. In Section 3 the data are described. Numerical results and discussion are presented in Section 4, followed by a brief conclusion.

**2- Minimal spanning tree and hierarchical tree construction**

Since the construction of the minimal spanning tree (MST) and hierarchical tree (HT) has been well explained in Mantegna [7], and Mantegna and Stanley [8], we will briefly present the methodology here. First, we define a correlation function between a pair of currencies based on the Exchange Rate Service [4] in order to quantify synchronization between the currencies. Let $P_i(t)$ be the rate $i$ at time $t$. Then, a rate change at a time interval $\tau$, $R_i(t)$, is defined as

$$R_i(t) = \ln P_i(t + \tau) - \ln P_i(t), \qquad (1)$$

meaning the geometrical change of $P_i(t)$ during the interval $\tau$. We take $\tau$ as one day in the following analysis throughout this paper. $R_i(t)$ is also called the vector of the time series of log-returns [22, 30]. Using the rate of change, the Pearson correlation coefficient between a pair of rates or individual currencies can be calculated by the cross-correlation function as

$$C_{ij} = \frac{\langle R_i R_j \rangle - \langle R_i \rangle \langle R_j \rangle}{\sqrt{\left(\langle R_i^2 \rangle - \langle R_i \rangle^2\right)\left(\langle R_j^2 \rangle - \langle R_j \rangle^2\right)}}, \qquad (2)$$

where $\langle ... \rangle$ represents the statistical average over the period studied. The correlation coefficient $C_{ij}$ have values ranging from -1 to +1, where -1 and +1 mean that two currencies, $i$ and $j$, are completely anti-correlated and correlated, respectively. If $C_{ij} = 0$ the currencies $i$ and $j$ are uncorrelated. The coefficients $C_{ij}$ form a symmetric $N \times N$ matrix with diagonal elements equal to unity. We apply the correlation matrix to construct a currency minimal spanning tree, and by using it we can intuitively understand the network among foreign exchange rates. The MST, a theoretical concept of graph theory [31], forms taxonomy for a topological space of the N rates. The MST is generated from a graph by selecting the most important correlations between currency indices. The MST reduces the information space from $N(N − 1)/2$ separate correlation coefficients to $(N − 1)$ linkages, known as tree ''edges'', while retaining the salient features of the system [23]. Therefore, the MST is a tree having $N - 1$ edges that minimize the sum of the edge distances in a connected weighted graph of the N rates. The edge distances satisfy the following three axioms of Euclidean distance:

(i)   $d_{ij} = 0$ if and only if $i = j$,
(ii)  $d_{ij} = d_{ji}$,
(iii) $d_{ij} \leq d_{ik} + d_{kj}$.

Here, $d_{ij}$ expresses the distance between each pair of currencies $i$ and $j$. We need the Euclidean distance between the currencies in order to construct the MST. However, it is well known that the correlation coefficient $C_{ij}$ does not satisfy all these axioms. One can convert the correlation coefficient by appropriate functions so that the axioms can be applied. One of the appropriate functions that fulfills the three axioms was found by Mantegna [7], and Mantegna and Stanley [8] and is defined as

$$d_{ij} = \sqrt{2(1 - C_{ij})}, \qquad (3)$$

where the distance $d_{ij}$ can lie in $0 \leq d_{ij} \leq 2$, while correlations run from -1 to +1. High correlations correspond to small values of $d_{ij}$.

Now, one can construct a MST for a pair of currencies using the $N \times N$ matrix of $d_{ij}$. Two methods for constructing the MST are Kruskal's algorithm [31-33] and Prim's algorithm [34]. We used Kruskal's algorithm [31-33] to construct our MST. The algorithm consists of the following steps: (i) we choose a pair of currencies with nearest distance and connect them with a line proportional to the distance; (ii) we also connect a pair of currencies with the 2nd nearest distance; (iii) we also connect the nearest pair that is not connected by the same tree. We repeat the third step until all the given rates are connected in one unique tree. Then, we achieve a connected graph without cycles.

Finally, in order to construct the HT, we introduce the ultrametric distance or the maximal $\hat{d}_{ij}$ between two successive currencies encountered when moving from the starting currency $i$ to the ending currency $j$ over the shortest part of the MST connecting the two currencies. For example, the distance $\hat{d}_{ad}$ is $d_{bc}$ when the MST is given as

$$a - b - c - d$$

where $\hat{d}_{bc} \geq max\{d_{ab}, d_{cd}\}$. The distance $\hat{d}_{ij}$ satisfies the axioms of Euclidean distance and the following ultrametric inequality with a condition stronger than axiom (iv) $\hat{d}_{ij} \leq \hat{d}_{ik} + \hat{d}_{kj}$ [35], which can be written as

$$\hat{d}_{ij} \geq max\{d_{ik}, d_{kj}\}. \qquad (4)$$

The distance $\hat{d}_{ij}$ is called the subdominant ultrametric distance [36, 37]. One can construct an HT by using Eq. (4).

We performed a technique to associate a value of reliability to the links of correlation-based graphs, namely the minimal spanning trees (MSTs) and hierarchical trees (HTs), by using bootstrap replicas of data. We also used the average linkage cluster analysis for obtaining the hierarchical trees in the case of the TL as the numeraire. This analysis gives the cluster structure much better. The bootstrap technique and average linkage cluster analysis were explained extensively in Refs. [5, 6].

The constructions of the MST and HT will be elaborated in Section 4.

## 3. Data

We chose thirty-four major currencies because they were generally free floating and had either market dominance or represented a region covering the data period from January 2, 2007 to December 31, 2008, as shown in Table 1. Data were provided from the Pacific Exchange Rate Service available online (http://fx.sauder.ubc.ca/data.html) which coincides with daily data. These data give some idea of how international currencies interact, how currency nodes are clustered, and the pattern behind price influences. We should also mention that these data represent a small sample compared to those of financial market studies; hence they will restrict possible topologies. We will construct the MSTs and the HTs from these data in the next section.

## 4. Numerical results and discussions

In this section, we present the MSTs and the HTs using the 34 major currencies which were studied to investigate the topology of the correlation networks among these currencies. The MSTs were constructed using Kruskal's algorithm [31-33] for the USD and the TL based on a distance-metric matrix. The reason of choosing the USD and the TL as numeraires is that the USD is the major currency and the TL is the minor currency. The numbers on the links of the MSTs and HTs are obtained from the bootstrap technique and the number gives information about statistical reliability of each link of graphs. If the values are close to one, the statistical reliability or the strength of the link is very high. Otherwise, the statistical reliability or the strength of the link is lower.

Figs. 1(a), 2(a) and 3(a) show the MSTs applying the method of Mantegna and Stanley [7, 8] for the USD based on a distance-metric matrix for the full years 2007, 2008 and for the 2007-2008 periods, respectively. The number of links in the MSTs is 32; hence the number of links should be (N-1); in our data, N=33, because we used 34 currencies in which one of them is a numeraire. From these trees, we detected different clusters of countries according to their geographical proximity and economic ties. In Fig 1(a), the first cluster is the Europe cluster that we focused on the EUR (Euro). All neighbors of the EUR are almost all European currencies, such as the CHF (Swiss Franc), CZK (Czech Koruna), GBP (British Pound), SEK (Swedish Krona), RUB (Russian Ruble) and NOK (Norwegian Kroner) along with the JPY (Japanese Yen). This grouping forms a set of currencies that are highly correlated with each other with the EUR at its center that the bootstrap values of the links between these currencies are changed from 0.82 to 1.0 (in a scale from 0 to1). The exception in this group is the JPY, which do not fit into the Asian-Pacific cluster but is generally linked to Western markets. The similar behavior with the EUR at the center was also reported in [1, 2, 38]. Although CNY (Chinese Renminbi) linked to the EUR, the bootstrap value of the link between EUR and CNY is 0.04. This means that this link is only a statistical fluctuation; hence the CNY does not belong to the Europe cluster. Similarly, from Fig. 1(a) the SAR (Saudi Arabian Riyal) and KWD (Kuwaiti Dinar) pair is observed as subgroup of the Europe cluster, but has no statistical justification according to the bootstrap analysis. Therefore, the KWD and SAR do not also belong to Europe cluster. The second cluster seems to the Asia-Pacific cluster which is composed of the Asia and Pacific countries, but the bootstrap analysis does not support this result, therefore one cannot consider as a cluster. We should also mention that the AUD (Australian Dollar) and NZD (New Zealand Dollar) are strongly connected with each other and similar results have been reported in Refs. [3, 22]. Moreover, the IDR (Indonesian Rupiah) and PHP (Philippines Peso) are also closely connected with each other in this group. It is expected that the THB (Thai Baht) and INR (Indian Rupee) to be in the Asia-Pacific group because these are Asian countries, but the THB and INR are not in this group. In this case, trade relationships and economic ties appear to dominate geographic effects. The South American cluster is a third cluster which is composed of the BRL (Brazilian Real), COP (Colombian Peso) and ARS (Argentinean Pesos). Although the EGP (Egyptian Pound) linked to the ARS, the bootstrap value of the link between the EGP and ARS is 0.02 and this means that this link is only a statistical fluctuation; hence the EGP does not belong to the South American cluster. This cluster linked to the Asia-Pacific cluster via the NZD and is also linked to the Europe cluster via the TL. Here, we can see that the TL plays a bridging role between the European and South American groups which the BRL heads. The connection between the BRL and TL was also reported by Brida et al. [11]. From Fig. 2(a), we can observe a cluster similar to that seen in Fig. 1(a), but following differences have been found. i) The Asia and the Pacific clusters are separated from each other. ii) The Asia grouping with the SGD (Singapore Dollar) at its head which is directly linked to the Europe grouping. The MST in Fig. 3(a) is very similar to Fig. 2(a) in terms of overall structure but the INR, PHP, IDR and EGP are split out from the Asian group and the INR and TL are strongly correlated with each other. Moreover, we also obtained the average bootstrap value of the links between the currencies and found that 0.6225, 0.5925 and 0.6012 for 2007, 2008 and, of the 2007-2008 periods, respectively. It is also worthwhile mentioning that the average bootstrap value for 2007 is larger than for 2008 and for 2007-2008 periods. This fact showed that when we used the USD as numeraire, the most stable tree is obtained for the full year of 2007. This is not surprising because of the foreign exchange markets have more stable structures before the global financial crisis. Finally, from these results, one can conclude that the global financial crisis had a serious effect on markets.

The HTs of the subdominant ultrametric space associated with the MST are shown in Figs. 1(b), 2(b) and 3(b). Two currencies (lines) link when a horizontal line is drawn between two vertical lines. The height of the horizontal line indicates the ultrametric distance at which the two currencies are joined. To begin with, in Fig 1(b), one can see that the distance between the SAR and KWD is the smallest of the sample, indicating a strong relationship between these two currencies which are form the first cluster. The second cluster is Europe which consists of two sub-groups, one with the HUF (Hungarian Forint) and PLN (Polish Zloty) and the other with the EUR and RUB. The third cluster, composed of the BRL, ZAR (South African Rand) and TL are related with each other when we choose the USD as the numeraire. One explanation is that these are currencies belonging to developing countries. In Fig. 2(b), one can see that the overall structure observed in this figure is consistent with Fig. 1(b), but the following relevant changes are detected: (i) there are three sub-groups in the Europe cluster in Fig. 2(b), in contrast to Fig. 1(b). These sub-groups are the CZK and PLN, the EUR and SKK (Slovakian Koruna), and the NOK and SEK (Swedish Krona). (ii) The Asia cluster made up of the PHP and INR. (iii) South America is an inhomogeneous cluster which consists of the BRL, MXN (Mexican Peso), ZAR and TL. Also in this cluster we can see two sub-groups which are the BRL and MXN, and the ZAR and TL. Similarly, from Fig. 3(b), we can see that the overall structure observed in this figure is consistent with Fig. 2(b), but the following relevant changes are detected. (i) The ZAR and TL leave the South American cluster and form a new independent cluster. (ii) The BRL and MXN stay with each other and form another independent cluster. (iii) The KRW (South Korean Won) and TWD (Taiwan Dollar) form a new cluster. Finally, the obtained bootstrap values of the links between the currencies consistent with these clusters which seen in Figs. 1(b), 2(b) and 3(b).

The MSTs shown in Figs. 4(a), 5(a) and 6(a) were obtained for the TL based on the distance-metric matrix for the full years of 2007, 2008 and for the 2007-2008 periods, respectively. The TL based MSTs, Figs. 4(a), 5(a) and 6(a), illustrate less clustering than when using the USD as numeraire, when comparing these figures with Figs. 1(a), 2(a) and 3(a). In analyzing Fig. 4(a), we can see that the USD and EUR are the predominant world currencies and we can see also that the two strongest economic clusters are the international cluster with the USD at its center and the Europe cluster with the EUR at its center. A similar finding was also obtained in Refs. [1, 2, 38]. These clusters which we studied appear to be organized principally according to a geographical criterion and economic relations. International grouping forms a set of currencies that are highly correlated with each other. There are also four pairs of currencies namely the SAR and KWD, the USD and CNY, the AUD and NZD, and the EUR and CHF in which the bootstrap values of the links between these currencies are very high seen in Figs. 4(a), 6(a); hence these currencies are very close to each other. Moreover, the bootstrap values of the links between the pairs of the SAR and KWD, the USD and CNY, the AUD and NZD and the EUR and SKK are also high seen in Fig. 5(a). The Europe cluster consists of European countries along with the BRL, ZAR, AUD and NZD. Within this cluster the bootstrap values of the links between some currencies are very high. For instance, the bootstrap values of the links between the EUR and CHF, and EUR and CZK are equal to 1.00 or the bootstrap values of the links between the EUR and SEK, and the EUR and NOK equal to 0.96 and 0.98, respectively. These high bootstrap values of the links show that there are strong correlations between these currencies. In addition, the Europe cluster is linked to the international cluster via the RUB as seen in Fig. 4(a). In Fig. 5(a), we observed similar clusters to those seen in Fig. 4(a), except that Asian currencies are not in the international cluster and these currencies are not formed a cluster due to the bootstrap analysis. While Fig. 6(a) has the same characteristics as Fig. 5(a), it has the following differences. i) The INR, KRW, PHP and TWD are not inside the Asia countries and they are connected to the international cluster with the USD at its center. ii) The bootstrap values of the links between JPY and USD, PHP and TWD, and MXN and ARS are 0.18, 0.12 and 0.12 respectively; hence these links are a statistical fluctuation. Therefore, the JPY, PHP, MXN and BRL do not belong to the international cluster. Similar situation is also seen in Europe cluster for the CAD (Canadian Dollar) NZD and AUD. Moreover, we also obtained the average bootstrap value of the links between the currencies and found that 0.6678, 0.5503 and 0.5815 for 2007, 2008 and of the 2007-2008 periods, respectively. We should also mention that the average bootstrap value for 2007 is larger than for 2008 and for 2007-2008 periods. This fact showed that when we used the TL as numeraire, the most stable tree is obtained for the full year of 2007. From the results of the TL as numeraire, we can conclude that either currency are all linked by common economic factors or currency traders pay more attention to the USD and EUR movements than to local factors. The similar results were also found in [3].

The HTs of the subdominant ultrametric space associated with the MSTs are shown in Figs. 4(b), 5(b) and 6(b). In Figs. 4(b), 5(b) and 6(b), one can see that the distance between the USD and CNY is the smallest of the sample, indicating a strong relationship between these two currencies which form the first cluster. This fact is also seen in Figs 4(a), 5(a), 6(a), namely the bootstrap values are equal to 1.00 in Fig. 4(a), 6(a) and 0.98 in Fig. 5(a). We also see two other clusters, namely the clusters between the SAR and KWD, and the AUD and NZD, in Figs. 4(b), 5(b), 6(b). On the other hand, the EUR and CHF form a cluster with each other in Figs. 4(b) and 6(b), but in Fig. 5(b), the EUR forms a cluster with SKK. These clusters seen in Figs. 4(b), 5(b) and 6(b) are consistent with similar clusters seen in Figs. 4(a), 5(a) and 6(a). This fact is also observed from bootstrap values of the links between these currency pairs.

From Figs. 1(a) - 6(a) and 1(b) - 6(b), both of the USD and the TL based on the distance-metric matrixes, we can see that the bootstrap values of the links between the SAR and KWD and the AUD and NZD are very high (almost equal to 1.00 ). These high bootstrap values of the links show that there are strong correlations between these currencies. One possible explanation for this situation is the dominance of economic ties (especially the oil trade) and geographical proximity for pairs of the SAR and KWD and the AUD and NZD. Another striking feature is the behavior of the THB. The bootstrap value of the link between the THB and other currencies is small in all MSTs. This fact show that the THB is not belongs to any cluster in MSTs. Unfortunately the cluster structure is not observed clearly in these figures.

In order to observe the cluster structure much better we used the average linkage cluster analysis in the HTs for the case of the TL as the numeraire, seen in Fig. 7. Figs. 7(a), 7(b), 7(c) are obtained for the full years of 2007, 2008 and for the 2007-2008 periods, respectively. From these figures, we observe the four clusters. The first strong cluster is the cluster with the USD and EUR are at its center. The JPY heads the second cluster that consists of the SEK, MXN, GBP, CLP (Chilean Peso), INR, PLN and NOK. Third cluster consists of the PHP, SKK, HUF, IDR and CAD with the SKK at its center. Finally, the AUD heads the fourth cluster that includes the NZD, THB, COP, ZAR and BRL. Moreover, the correlation between the USD and CNY is the strongest and the second strong relationship is between the SAR and KWD. From these results and as well as Figs. 4(b), 5(b), 6(b), we can conclude that if the minor currency (TL) is used as the numeraire, the effect of the global financial crisis is not as obvious. Therefore, in order to see the effect of the global financial crisis one should take the major currency as choosing the "right" numeraire.

In conclusion, in this study, we presented the topology of correlation networks among 34 major currencies using the concept of the MST and the HT for the full years of 2007, 2008 and for the 2007-2008 periods when major economic turbulence occurred. The clustered structure of the currencies and the key currency in each cluster were obtained, and we found that the clusters matched nicely with geographical regions of corresponding countries in the world such as Asia or Europe. We performed a technique to associate a value of reliability to the links of MSTs and HTs by using bootstrap replicas of data to obtain the information about reliability of each link of graphs. From results of the bootstrap analysis, we can see that in general, the bootstrap values in MSTs and HTs are highly consistent with each other. We also obtained the average bootstrap value of the links between the currencies and found that 0.6225, 0.5925 and 0.6012 for the USD as the numeraire, and 0.6678, 0.5503 and 0.5815 for the TL as the numeraire for the full years of 2007, 2008 and, for the 2007-2008 periods, respectively. In both cases, we can see that the average bootstrap value for 2007 is larger than for 2008 and for 2007-2008 periods; hence the most stable tree is obtained for the full year of 2007 that is the year before the global financial crisis. Moreover, one can conclude from these facts that the TL based trees were more affected than the USD based trees from the global financial crisis. As expected, our results confirm that the USD and EUR are the predominant world currencies and that they are also both linked to other currencies according to their geographic location and related factors such as economic ties. Our results also show that the global financial crisis had an overwhelming effect on exchange rate liaisons and commercial relations. These results imply that currencies are either all linked by common economic factors or currency traders pay more attention to the USD and EUR movements than to local factors. Moreover, the results of our analysis demonstrate that economic dependency can be clarified using the theory of correlation networks; hence, we expect that correlation networks will be helpful in future monetary system studies. We also used the average linkage cluster analysis for obtaining the hierarchical trees in the case of the TL as the numeraire to observe the cluster structure much better. Finally, we should also mention that the detected hierarchical structures might be useful in theoretical descriptions of currencies and in the search for economic factors affecting specific groups of countries.

**Acknowledgement**

One of us (MK) would like to thank M. Takayasu and H. Takayasu for many useful discussions and critical readings of the manuscript as well as for their hospitality in the Department of Computational Intelligence and Systems Science, Tokyo Institute of Technology for his short visit. This work was supported by the Scientific and Technological Research Council of Turkey (TÜBİTAK) Grant No: 109T133. One of us (B.D.) would like to express his gratitude to TÜBİTAK for his PhD scholarship. B. D. is also grateful to R. N. Mantegna and F. Lillo for very helpful discussions about the bootstrap technique and the average linkage cluster analysis.

**Figure Captions**

**Fig. 1. a)** USD-based minimal spanning tree for the year 2007. Graphical representation of minimal distance metrics for currencies quoted against the USD.

**b)** USD-based hierarchical tree of subdominant ultrametric space for the year 2007. Hierarchical grouping of distance metrics for currencies quoted against the USD.

**Fig. 2. a)** Same as Fig. 1a, but for the year 2008

**b)** Same as Fig. 1b, but for the year 2008

**Fig. 3. a)** Same as Fig. 1a, but for the years 2007- 2008

    **b)** Same as Fig. 1b, but for the years 2007-2008

**Fig. 4. a)** TL-based minimal spanning tree for the year 2007. Graphical representation of minimal distance metrics for currencies quoted against the TL.

    **b)** TL-based hierarchical tree of subdominant ultrametric space for the year 2007. Hierarchical grouping of distance metrics for currencies quoted against the TL.

**Fig. 5. a)** Same as Fig. 4a, but for the year 2008

    **b)** Same as Fig. 4b, but for the year 2008

**Fig. 6. a)** Same as Fig. 4a, but for the years 2007- 2008

    **b)** Same as Fig. 4b, but for the years 2007-2008

**Fig. 7.** TL-based hierarchical tree by using the average linkage cluster analysis.
    **a)** For the year of 2007.
    **b)** For the year of 2008.
    **c)** For the years of 2007 - 2008.

**Table captions**

**Table 1.** Set of daily data for 34 currencies.

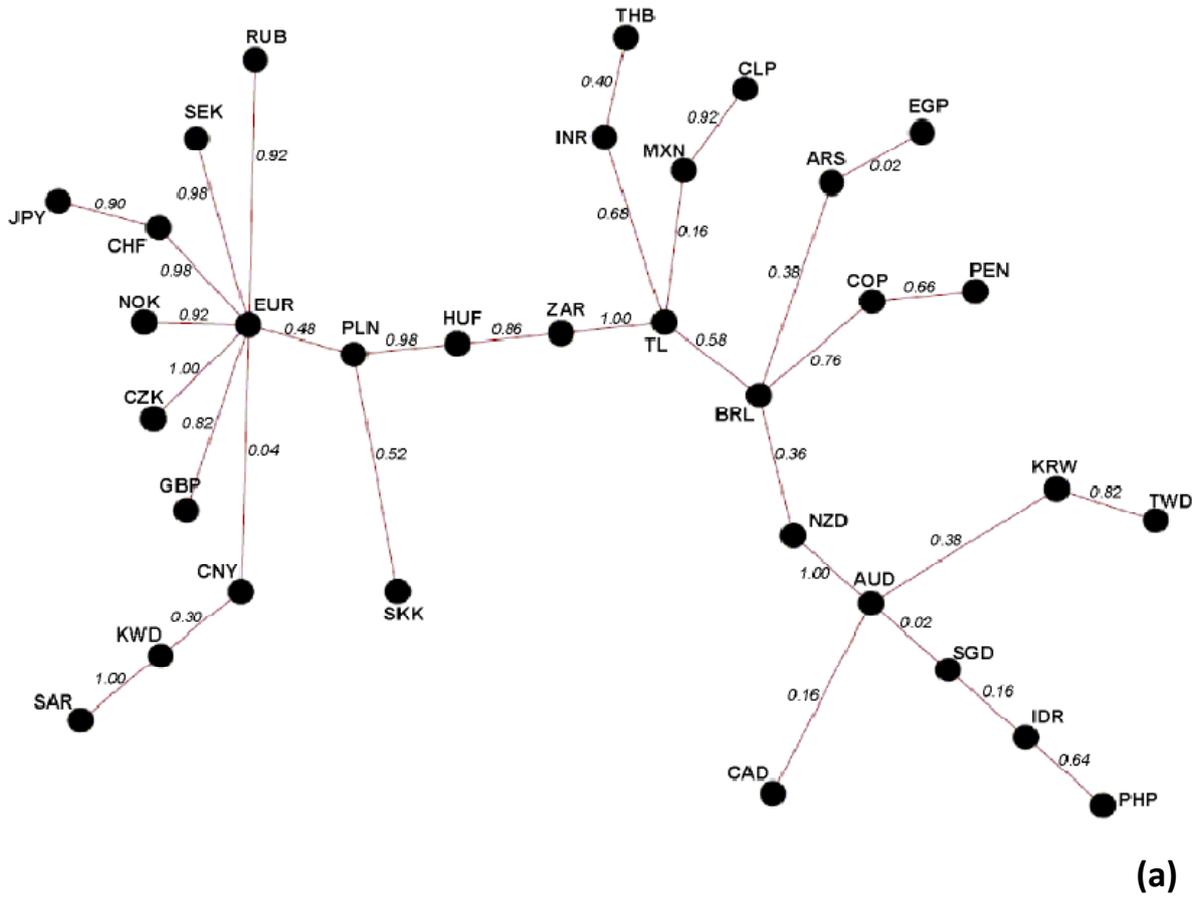

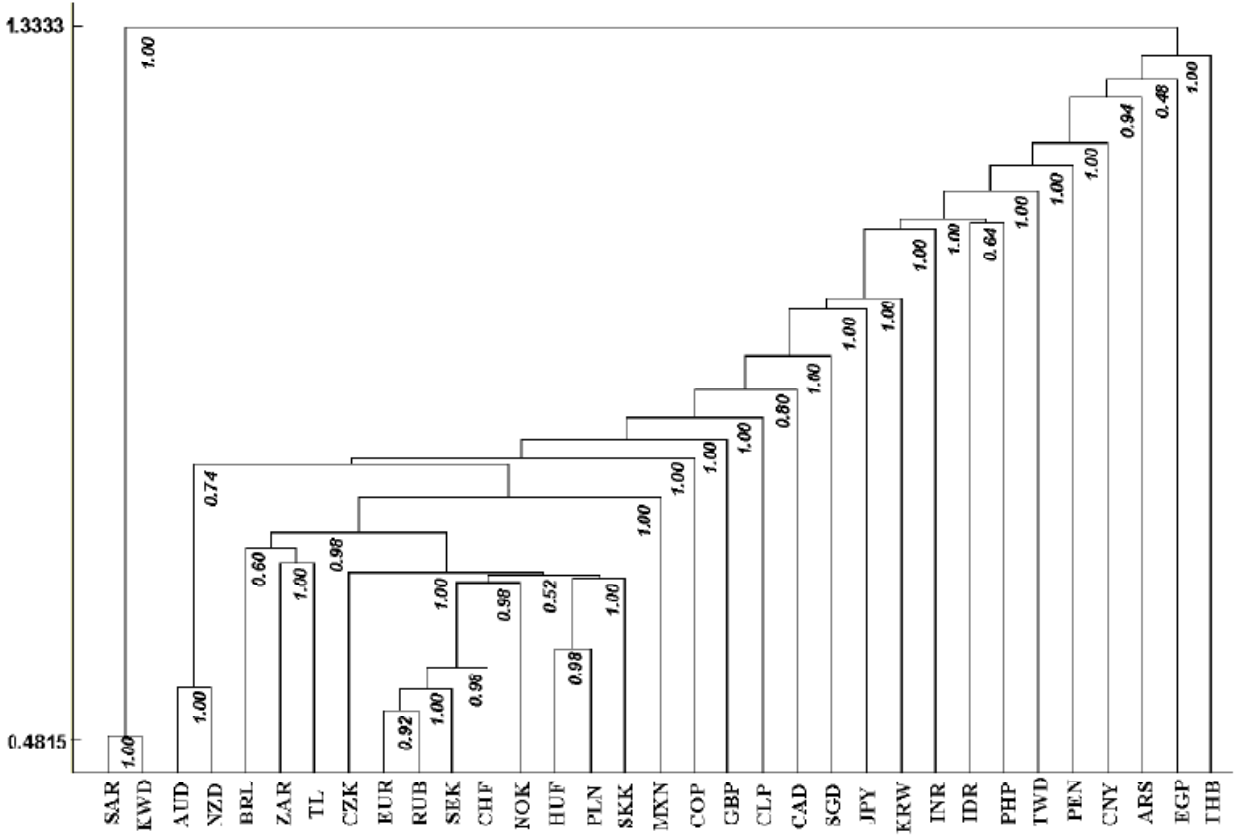

**Fig.1**

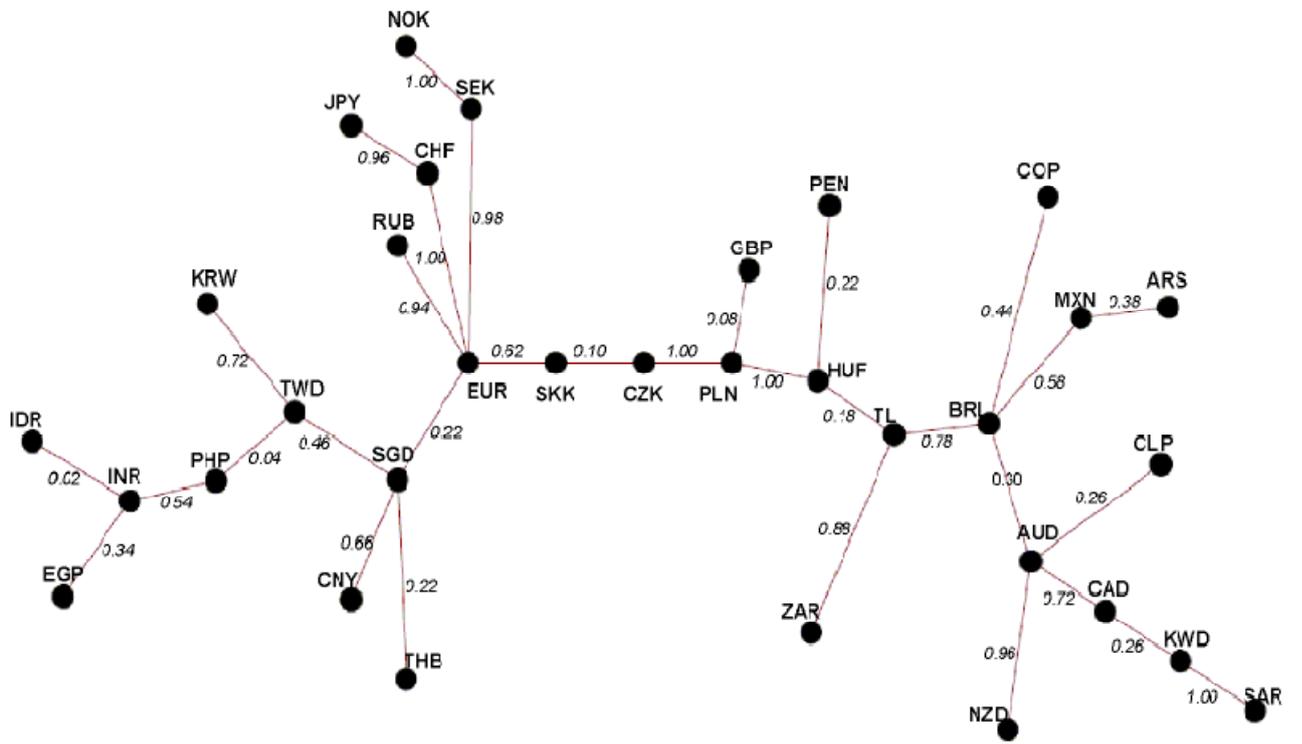
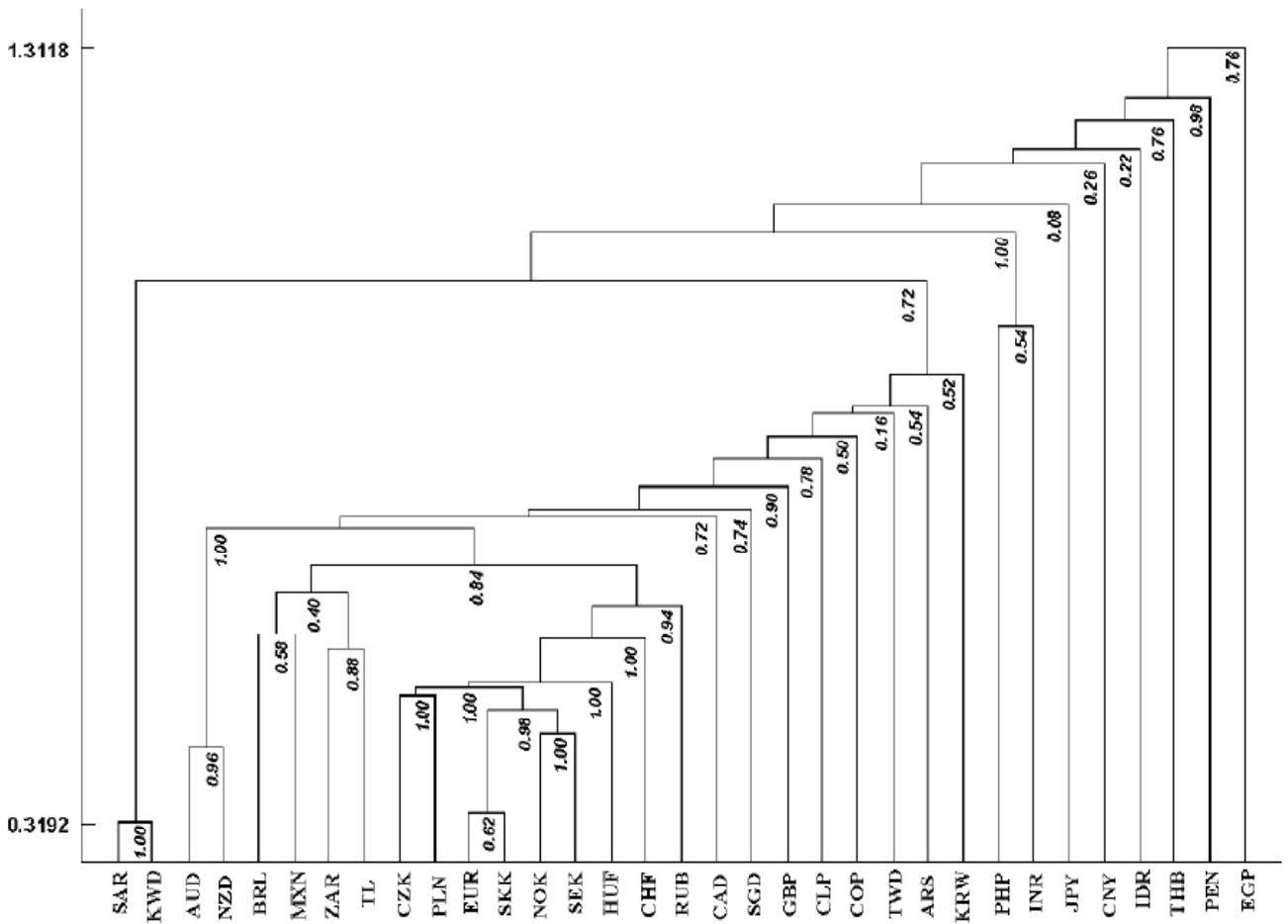

Fig.2

**Fig.3**

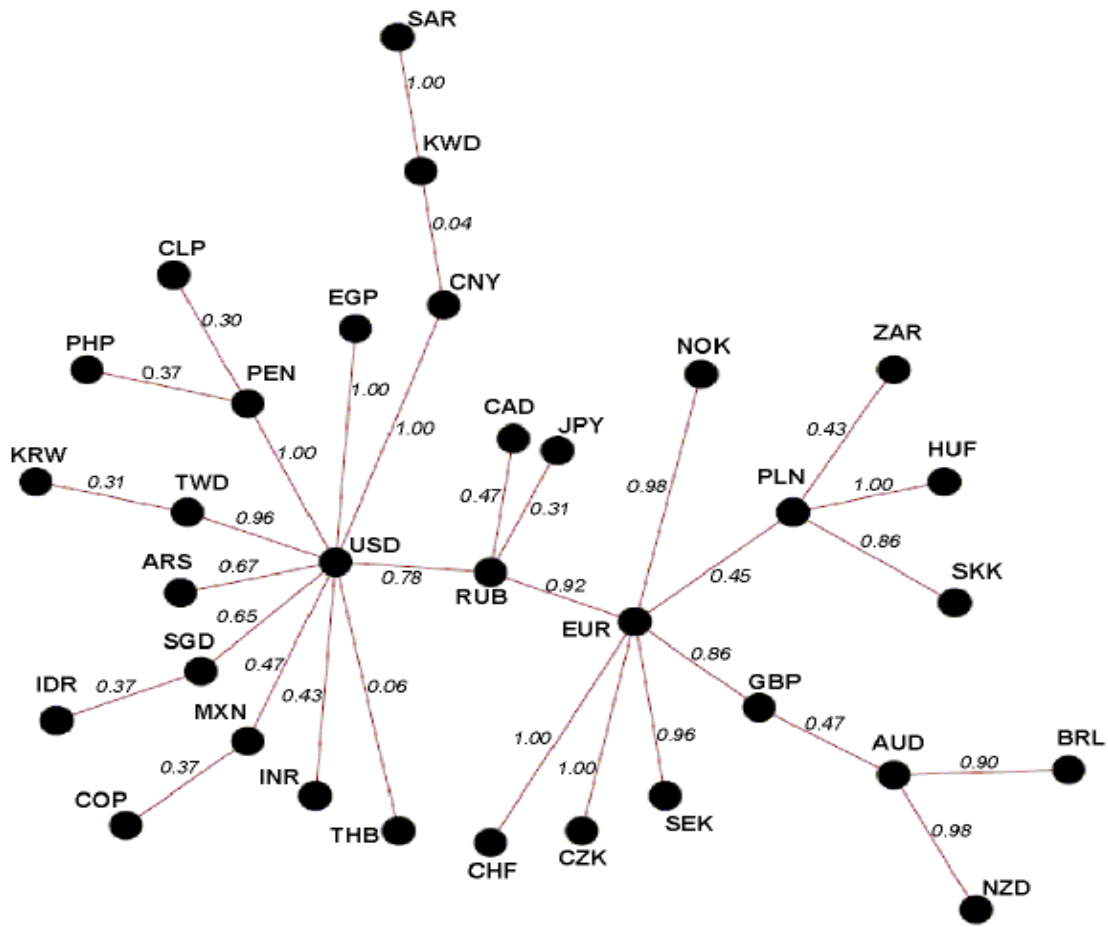

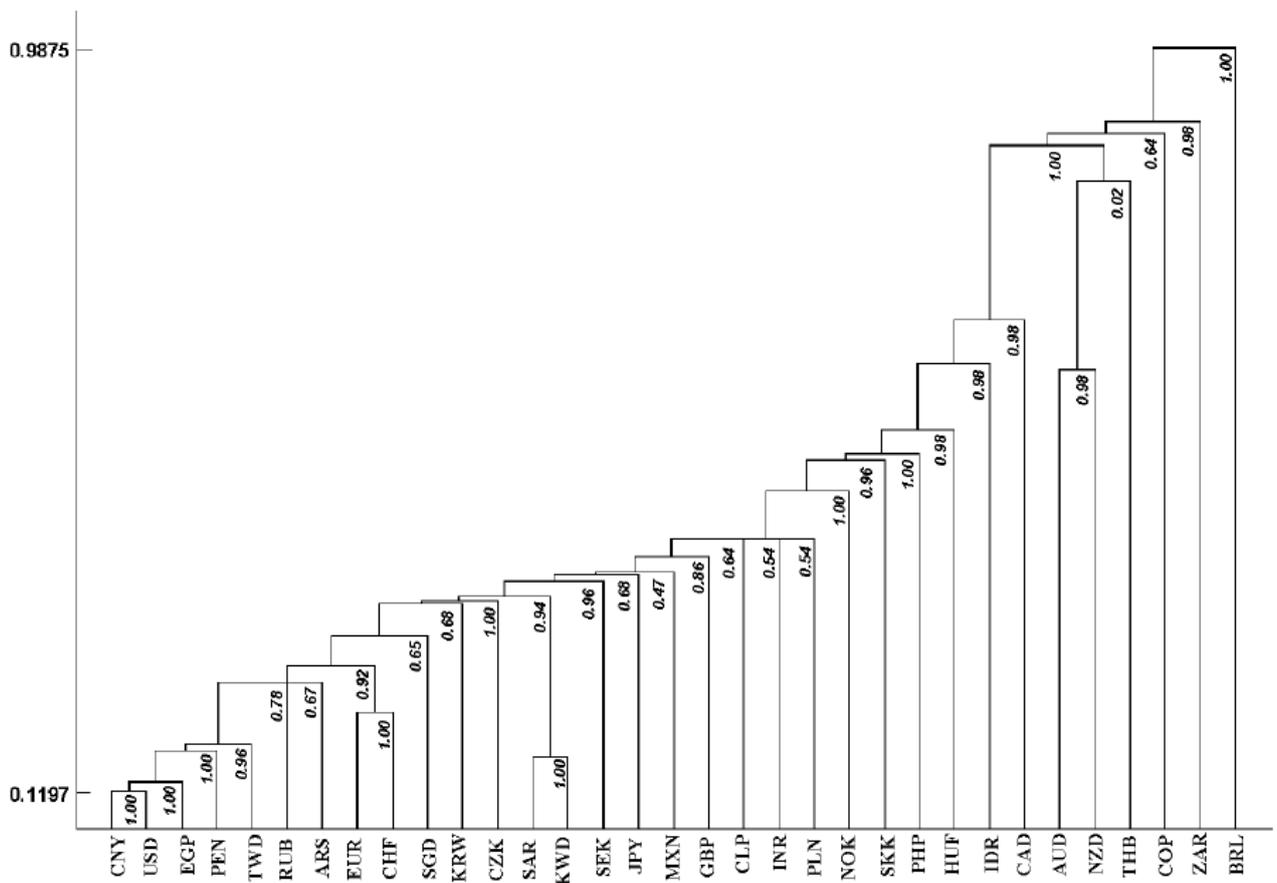

**Fig.4**

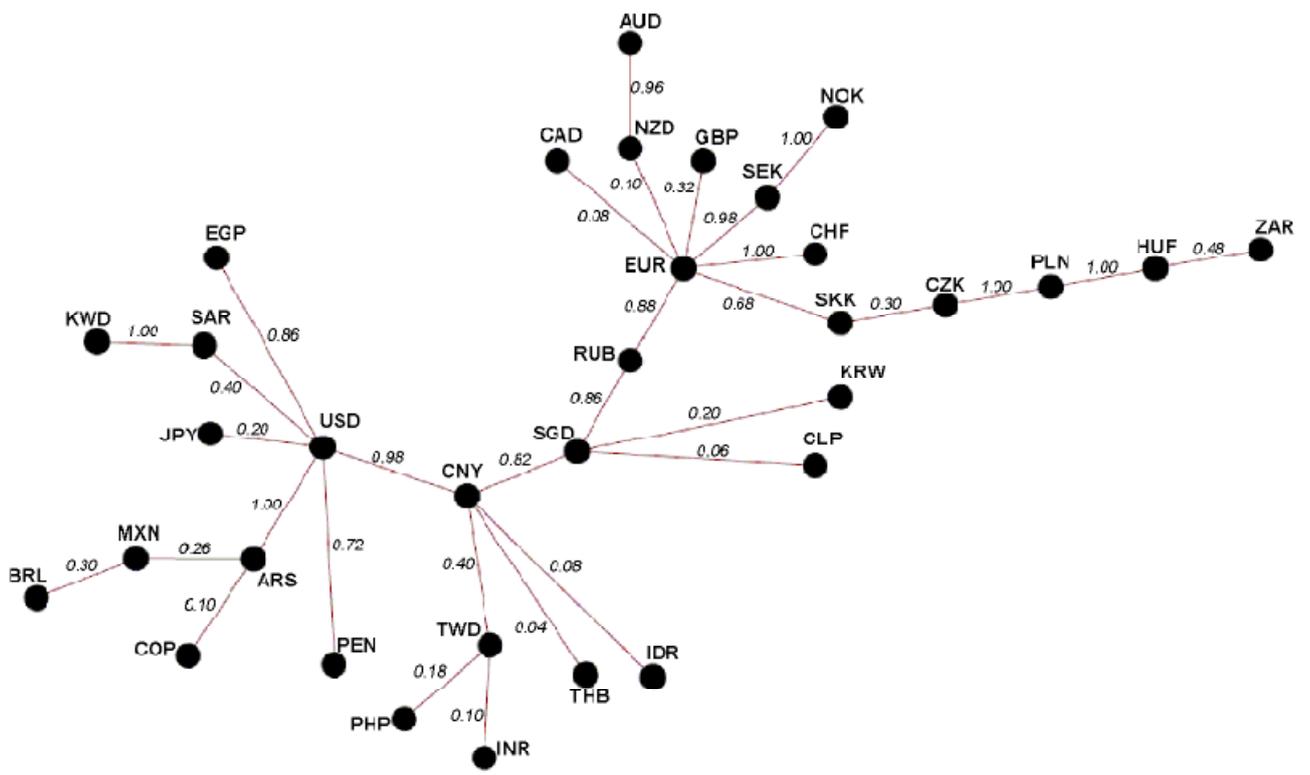

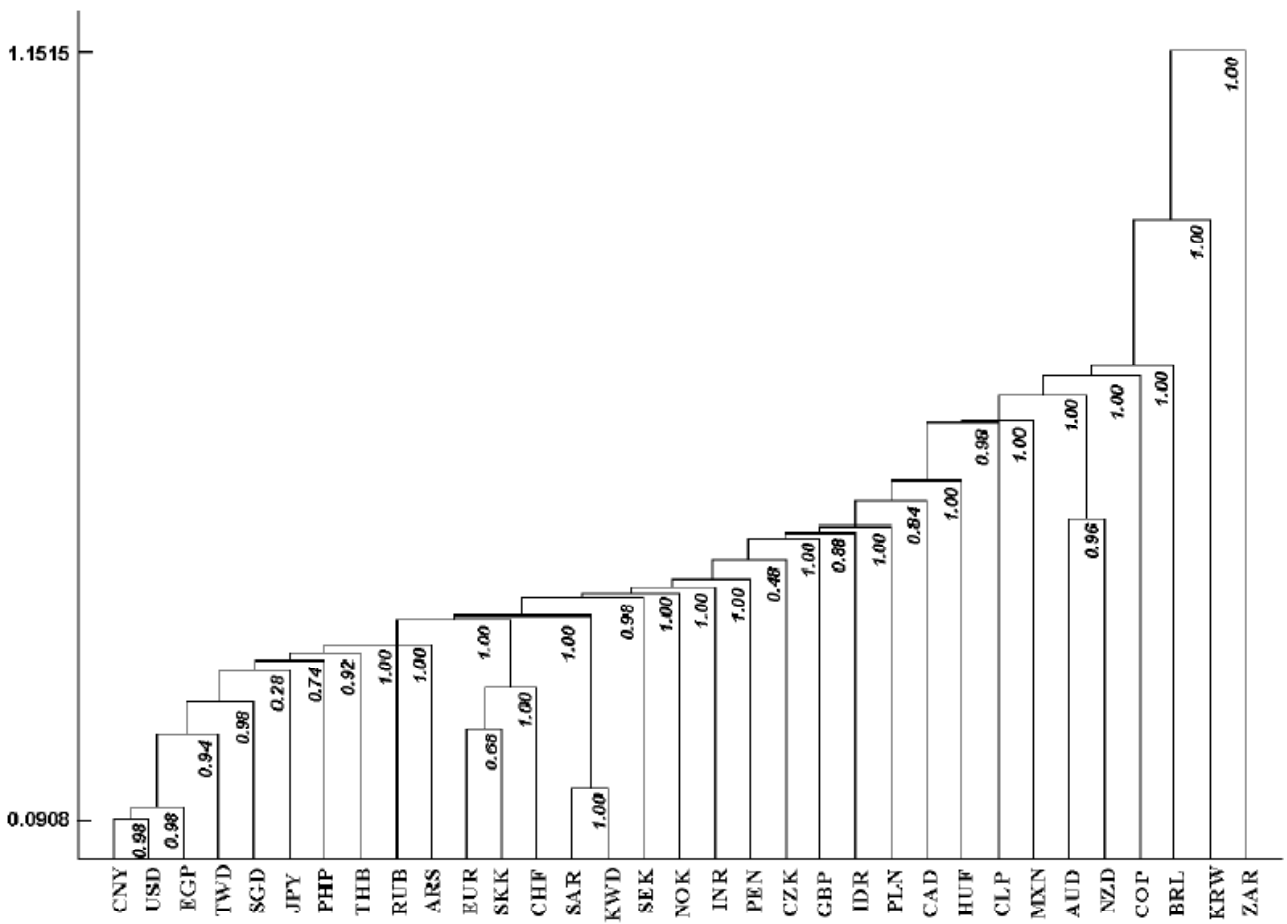

Fig.5

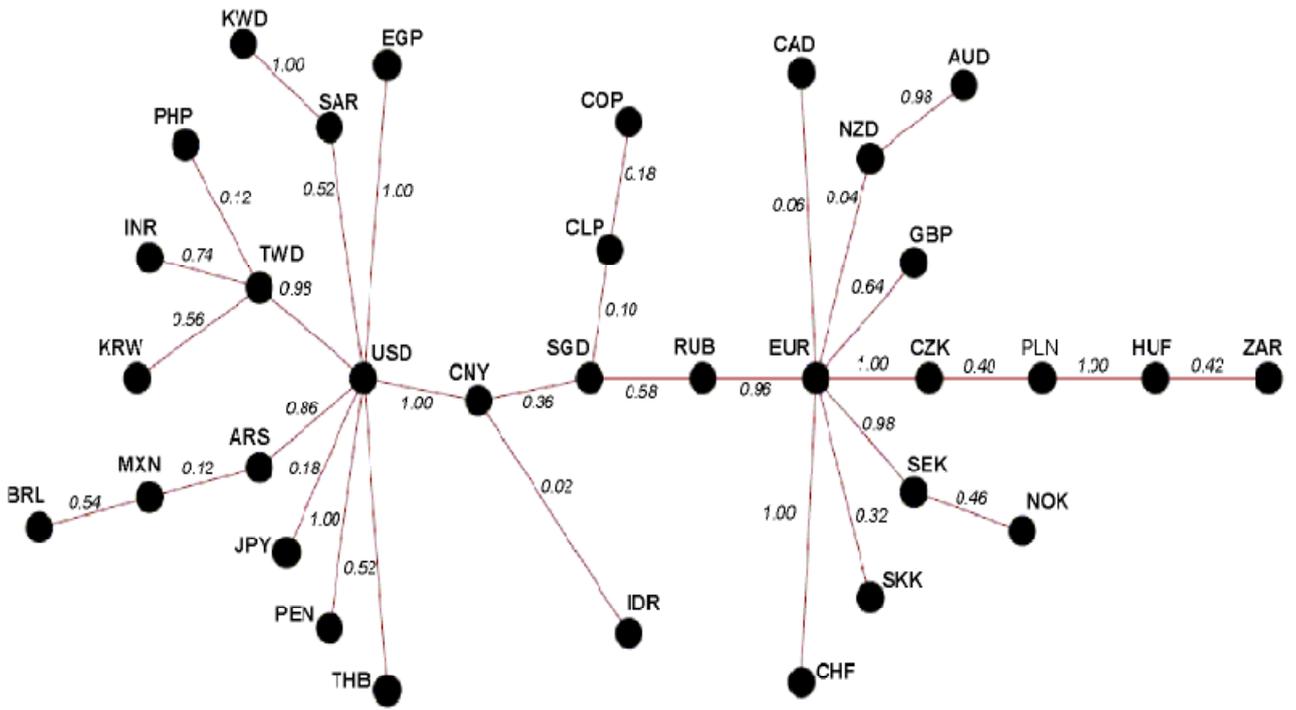
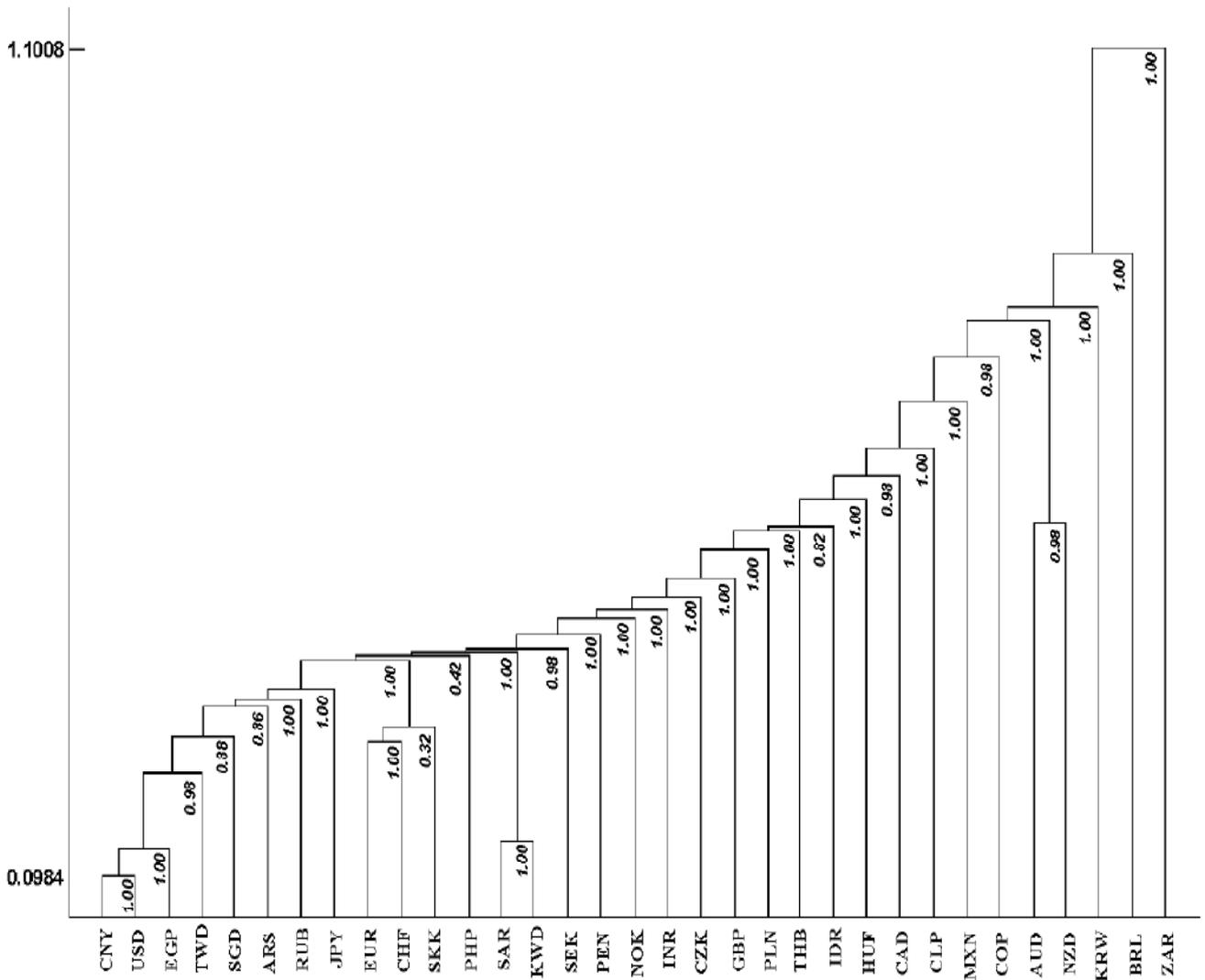

**Fig.6**

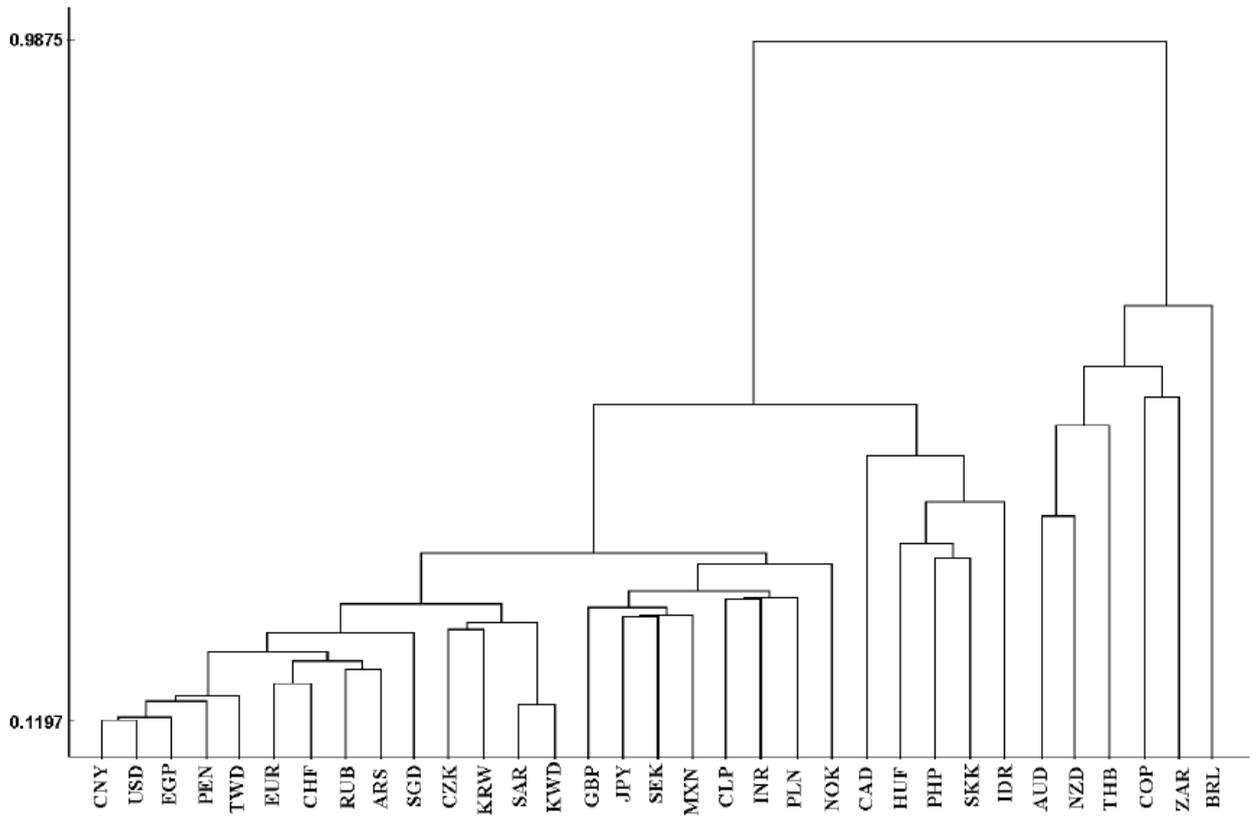

7 (a)

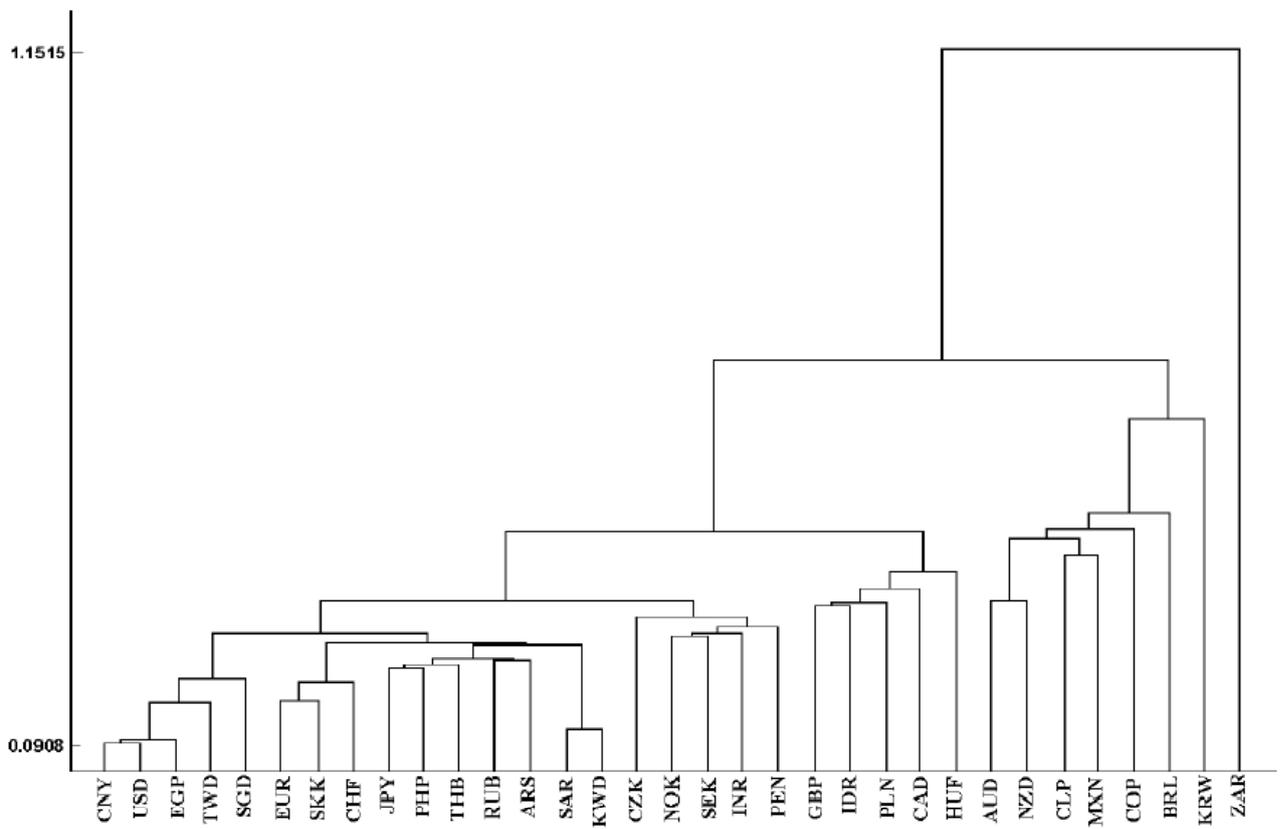

7 (b)

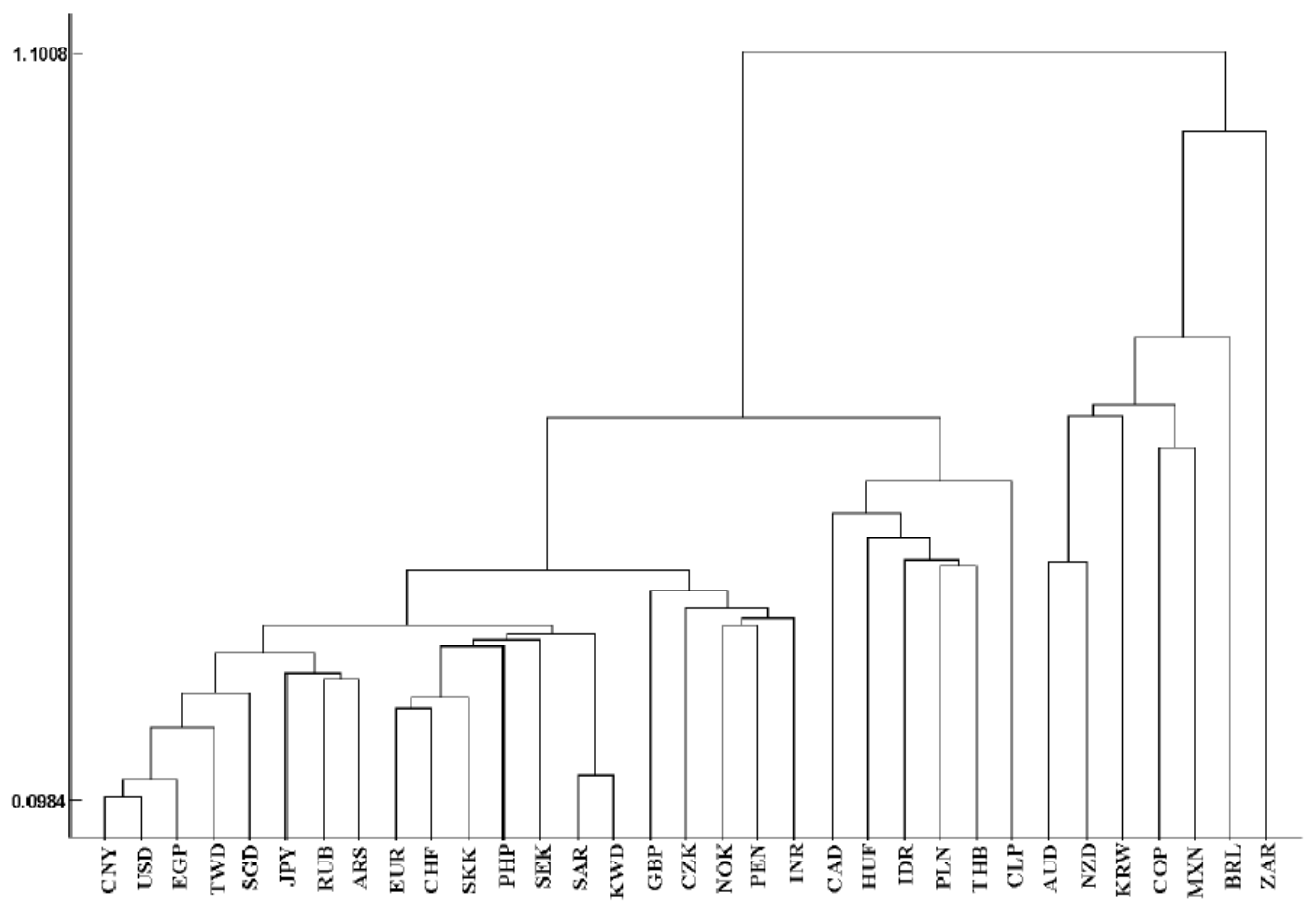

7 (c)

Fig.7

| Currency | Code | Currency | Code |
|---|---|---|---|
| Argentinian Pesos | ARS | New Zealand Dollar | NZD |
| Australian Dollar | AUD | Norwegian Krone | NOK |
| Brazilian real | BRL | Peruvian New Sole | PEN |
| British Pound | GBP | Philippines Peso | PHP |
| Canadian Dollar | CAD | Polish Zloty | PLN |
| Chilean Peso | CLP | Russian Ruble | RUB |
| Chinese Renminbi | CNY | Saudi Arabian Riyal | SAR |
| Colombian Peso | COP | Singapore Dollar | SGD |
| Czech Koruna | CZK | Slovakian Koruna | SKK |
| Egyptian pound | EGP | South African Rand | ZAR |
| Euro | EUR | South Korean Won | KRW |
| Hungarian Forint | HUF | Swedish Krona | SEK |
| Indian Rupee | INR | Swiss Franc | CHF |
| Indonesian Rupiah | IDR | Taiwan Dollar | TWD |
| Japanese Yen | JPY | Thai Baht | THB |
| Kuwaiti Dinar | KWD | Turkish Lira | TL |
| Mexican Peso | MXN | US Dollar | USD |

**Table 1**